\documentclass[twocolumn,english,twocolumn,prX]{revtex4}
\usepackage[T1]{fontenc}
\usepackage[latin9]{inputenc}
\usepackage{amsmath}
\usepackage{graphicx}
\usepackage{amssymb}

\makeatletter

\providecommand{\tabularnewline}{\\}

\@ifundefined{textcolor}{}
{%
 \definecolor{BLACK}{gray}{0}
 \definecolor{WHITE}{gray}{1}
 \definecolor{RED}{rgb}{1,0,0}
 \definecolor{GREEN}{rgb}{0,1,0}
 \definecolor{BLUE}{rgb}{0,0,1}
 \definecolor{CYAN}{cmyk}{1,0,0,0}
 \definecolor{MAGENTA}{cmyk}{0,1,0,0}
 \definecolor{YELLOW}{cmyk}{0,0,1,0}
 }

\makeatother

\makeatother

\usepackage{babel}

\makeatother

\usepackage{babel}

\begin{document}

\title{Why are nonlinear fits so challenging?}

\author{Mark K. Transtrum}

\affiliation{Laboratory of Atomic and Solid State Physics, Cornell University,
Ithaca, New York 14853, USA}

\email{mkt26@cornell.edu}

\author{Benjamin B. Machta}

\affiliation{Laboratory of Atomic and Solid State Physics, Cornell University,
Ithaca, New York 14853, USA}

\email{bbm7@cornell.edu}

\author{James P. Sethna}

\affiliation{Laboratory of Atomic and Solid State Physics, Cornell University,
Ithaca, New York 14853, USA}

\email{sethna@lassp.cornell.edu}
\begin{abstract}
Fitting model parameters to experimental data is a common yet often
challenging task, especially if the model contains many parameters.
Typically, algorithms get lost in regions of parameter space in which
the model is unresponsive to changes in parameters, and one is left
to make adjustments by hand. We explain this difficulty by interpreting
the fitting process as a generalized interpolation procedure. By considering
the manifold of all model predictions in data space, we find that
cross sections have a hierarchy of widths and are typically very narrow.
Algorithms become stuck as they move near the boundaries. We observe
that the model manifold, in addition to being tightly bounded, has
low extrinsic curvature, leading to the use of geodesics in the fitting
process. We improve the convergence of the Levenberg-Marquardt algorithm
by adding geodesic acceleration to the usual step.
\end{abstract}
\maketitle
The estimation of model parameters from experimental data is astonishingly
challenging. A nonlinear model with tens of parameters, fit (say)
by least-squares to experimental data, can take weeks of hand-fiddling
before a qualitatively reasonable agreement can be found; even then,
the parameters cannot usually reliably be extracted from the data.
Both general minimization algorithms and algorithms like Levenberg-Marquardt
that are designed for least-squares fits routinely get lost in parameter
space. This becomes a serious obstacle to progress when one is unsure
of the validity of the model, e.g.~in systems biology where one wants
to automatically generate and explore a variety of alternative models.

Here we use differential geometry to explain why fits are so hard.
We first explore the structure of the {\em model manifold} $\mathcal{M}$,
the manifold of predictions embedded in the space of data, $D$, and
find that it is typically bounded, with cross sections having a hierarchy
of widths, so that the overall structure is similar to that of a long,
thin ribbon. We explain this hierarchy by viewing the fitting process
as a generalized interpolation procedure with few effective model
degrees of freedom. We interpret the difficulty in fitting to be due
to algorithms getting stuck near the boundary of $\mathcal{M}$, where
the model is unresponsive to variations in the parameters. We then
discuss how geometry motivates algorithms to alleviate this difficulty.

\begin{figure}
\includegraphics[width=3.25in]{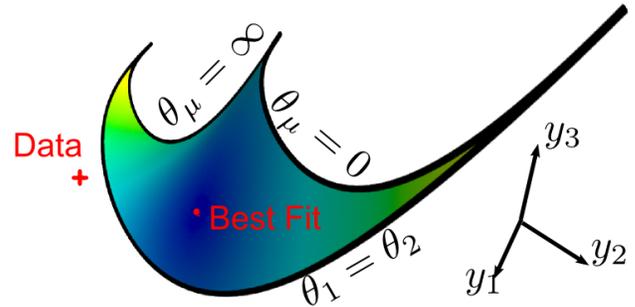}

\caption{\label{fig:ModelManifold} The model manifold for the two-exponential
problem, with $y_{i}$ evaluated at $t=1/3$, $1$, and $3$. Boundaries
exist when $\theta_{\mu}=0,\infty$ and when $\theta_{1}=\theta_{2}$.
(The ribbon-like structure of Fig.~\ref{fig:Needle} emerges only
in higher dimensions.)}

\end{figure}

A typical nonlinear least squares problem fits a model $Y_{m}(\theta)$
with $N$ parameters $\theta$ to $M$ experimental data points $y_{m}$.
We define the model manifold, $\mathcal{M}$, as the parametrized
$N$-dimensional surface $\vec{Y}(\theta)$ embedded in Euclidean
data space, $D=\mathbb{R}^{M}$. The best fit to the experiment is
given by the point on $\mathcal{M}$ with Euclidean distance closest
to the data, minimizing the cost $C=\frac{1}{2}(\vec{Y}(\theta)-\vec{y})^{2}$.
The Euclidean metric of data space (with distance between models given
by the change in residuals, $\vec{r}=\vec{Y}(\theta)-\vec{y}$) induces
a metric on the manifold, $g_{\mu\nu}=\partial_{\mu}\vec{Y}\cdot\partial_{\nu}\vec{Y}=(J^{T}J)_{\mu\nu}$,
where $J_{m\mu}=\frac{\partial}{\partial\theta_{\mu}}Y_{m}$; $g_{\mu\nu}$
is known as the Fisher Information matrix. As an example, the model
$Y(t,\theta)=f_{\theta}(t)=e^{-\theta_{1}t}+e^{-\theta_{2}t}$ sampled
at three time points is given in Fig.~\ref{fig:ModelManifold}. The
model manifold has been extensively studied by the information geometry
statistics community~\cite{Amari2007}, but they focus on the intrinsic
curvature; as the cost is the distance in data space, the embedding
and its extrinsic curvature are crucial to finding best fits~\cite{Bates1980,Bates1988}.

\begin{figure}
\includegraphics[width=3.25in]{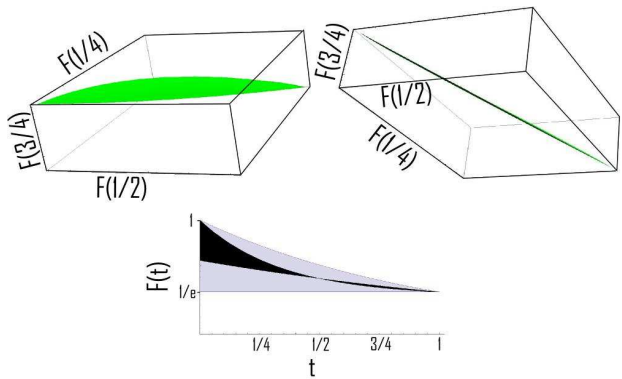}

\caption{\label{fig:Needle} Top: Two views of the cross section of the model
manifold for an {\em infinite} sum of exponentials $F_{{\mathbf{A}},{\mathbf{\theta}}}(t)=\sum_{n}A_{n}\exp(-\theta_{n}t)$
with $A_{n}\ge0$, given by fixing $F(0)=1$ and $F(1)=1/e$. Bottom:
The range of allowed fits (grey) is strongly reduced by fixing the
output at $t=1/2$ to the midpoint of its range (black).}

\end{figure}

\begin{figure}
\includegraphics[width=3.25in]{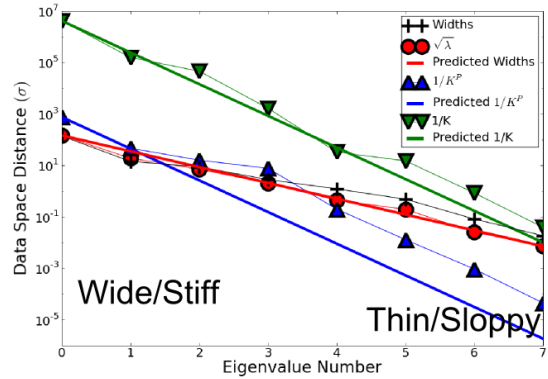} 

\caption{\textbf{\label{fig:Cross-sectional-widths}} Geodesic cross-sectional
widths of an eight dimensional model manifold along the eigendirections
of the metric from some central point, together with the square root
of the eigenvalues (singular values of the Jacobian), the inverse
extrinsic (geodesic) curvature $K$ \cite{Spivak1979}, and the inverse
geodesic parameter-effects curvature $K^{P}$ \cite{Bates1980,Bates1988,Transtrum2010}.
Notice the hierarchy of these data-space distances -- the widths and
singular values each spanning around four orders of magnitude and
the curvatures covering eight. Note also that the extrinsic curvatures
are three orders of magnitude smaller than the parameter-effects curvature.}

\end{figure}

As seen in figs.~\ref{fig:Needle} and~\ref{fig:Cross-sectional-widths},
this model manifold can take the form of a hyper-ribbon, with thinnest
direction four orders of magnitude smaller than the long axes. To
understand this observed hierarchy, consider the special case of analytic
models, $f(t,\theta)$, of a single independent variable (time) where
the data points are $Y_{m}=f(t_{m})$. Let $R$ be the typical time
scale over which the model behavior changes, so that the $n^{th}$
term of the Taylor series $f^{(n)}(t)/n!\lesssim R^{-n}$ (roughly
the radius of convergence). If the function is sampled at $n$ time
point $(t_{1},t_{2},...,t_{n})$ within this time scale, the Taylor
series may be approximated by the unique polynomial of degree $n-1$,
$P_{n-1}(t)$ passing through these points. At a new point, $t_{0}$,
the discrepancy between the interpolation and the function is given
by \begin{equation}
f(t_{0})-P_{n-1}(t_{0})=\omega_{n}(t_{0})f^{(n)}(\xi)/n!,\label{eq:RangeofBehavior}\end{equation}
 where $\xi$ lies somewhere in the interval containing $t_{0},\ t_{1},\ ...,\ t_{n}$~\cite{Stoer2002}.
The polynomial $\omega_{n}(t)$ has roots at each of the interpolating
points $\omega_{n}(t)=(t-t_{1})(t-t_{2})...(t-t_{n}).$ The possible
error of the interpolation function bounds the allowed range of behavior,
$\Delta f_{n}$, of the model at $t_{0}$ after constraining the nearby
$n$ data points (i.e. cross sections). Consider the ratio of successive
cross sections, $\frac{\Delta f_{n+1}}{\Delta f_{n}}=(t-t_{n+1})(n+1)\frac{f^{n+1}(\xi)}{f^{n}(\xi')}.$
If $n$ is sufficiently large, then $(n+1)\frac{f^{n+1}(\xi)}{f^{n}(\xi')}\approx\frac{1}{R}$;
therefore, we find that $\frac{\Delta f_{n+1}}{\Delta f_{n}}\approx\frac{t-t_{n+1}}{R}<1$
by the ratio test. Each cross section is thinner than the last by
a roughly constant factor, yielding the observed hierarchy.

We argue that this hyper-ribbon structure will be shared with a wide
variety of nonlinear, multiparameter models. Note that the eigenvalues
of the metric tensor $g_{\mu\nu}$ in Fig.~\ref{fig:Cross-sectional-widths}
also form a hierarchy, spanning eight orders of magnitude -- this
`sloppiness' has been documented in a number of other models, including
seventeen in systems biology~\cite{Gutenkunst2007a}, insect flight
and variational quantum wave functions~\cite{Waterfall2006}, interatomic
potentials~\cite{Frederiksen2004}, and a model of the next-generation
international linear collider~\cite{Gutenkunst2008}. (Multiparameter
models whose parameters are individually measured by the data, and
on the other extreme models with sensitive, chaotic dependence on
parameters, will likely not fall into this family.) Most parameters
in these models have bounded effects -- they can be set to limiting
values (zero, infinity, etc.) and still have finite model predictions
-- rates and Michaelis-Menten coefficients in systems biology, Jastrow
and determinential factors in variational wave functions, etc. Note
that the widths of the model manifold track nicely with these eigenvalues
in Fig.~\ref{fig:Cross-sectional-widths} (taking the square root
to match units): moving parameter combinations along eigendirections
of the metric by an order of magnitude (fixed shift in log parameters)
exhausts the range of behavior (width). This tracking suggests that
the ubiquity of sloppy eigenvalue spectra at best fits implies a ubiquitous
hyper-ribbon structure for the model manifold.

Our observation that many models are sloppy, presumably sharing this
hyper-ribbon model manifold structure, is now explained: multiparameter
models are a kind of high-dimensional analytic interpolation scheme,
and near degenerate Hessians result whenever multiple data points
reside within some generalized radius of convergence. When this is
the case, the data points are highly correlated and the model has
few effective degrees of freedom. Whenever there are many model parameters
for each effective degree of freedom there will be a hierarchy of
widths and the model will be sloppy.

Our geometric interpretation explains a number of observations about
nonlinear models. First, although parameters cannot be reliably extracted
by fitting degenerate models, it is still possible to constrain the
outcome of new experiments~\cite{Brown2003a,Gutenkunst2007a}. Because
the fitting process is an interpolation scheme, only a few stiff parameter
combinations need to be tuned to fit most of the data, since only
a few data points constrain the predictions at other times. The remaining
unconstrained parameter combinations control the interpolated values,
which are already restricted by the analyticity of the model.

Figure~\ref{fig:Cross-sectional-widths} also shows that the {\em
parameter effects curvature}~\cite{Bates1980,Bates1988,Transtrum2010}
and the geodesic extrinsic curvatures vary over twice as many decades
as the widths and $\sqrt{\lambda}$; indeed, their formulas include
a factor of $1/\lambda$%
\footnote{$K=\left\Vert P\partial_{\mu}\partial_{\nu}\vec{r}v^{\mu}v^{\nu}\right\Vert /v^{\alpha}v^{\beta}g_{\alpha\beta}$
where $P$ is a projection operator projecting out of the tangent
space for extrinsic curvature and into the tangent space for parameter
effects curvature. %
}. Why is the extrinsic curvature so much smaller? The manifold has
zero extrinsic curvature if there are equal numbers of parameters
as data points, $N=M$ (where the model manifold is a sub-volume in
the Euclidean data space). We have seen that most of the data points
are interpolations that supply little new information; the extrinsic
curvature will be small when the effective dimensionality of the embedding
space is not larger than than the number of parameters. %
\footnote{If we choose $N$ independent data points as our parametrization,
then the interpolating polynomial, $P_{N-1}(t)$ in Eq.~\ref{eq:RangeofBehavior}
is linear in the parameters. As discussed below that equation,
the manifold in each additional direction will be constrained to within
$\epsilon=\Delta f_{N+1}$ of $P_{N-1}(t)$. Presuming that this deviation
from flatness smoothly varies along the $j$th largest width $W_{j}\sim\Delta f_{j}$
(i.e., no complex or sensitive dependence on parameters),
the geodesic extrinsic curvature is $\epsilon/W_{j}^{2}$, explaining
the range of curvatures in Fig.~\ref{fig:Cross-sectional-widths}.
The ratio of the curvature to the inverse width should then be $\epsilon/W_{j}\sim\Delta f_{N+1}/\Delta f_{j}\sim(\Delta t/R)^{N+1-j}$,
comparable to the widths along the sloppiest direction and with twice
the slope. It appears that the scale of the parameter-effects curvature
is set by the stiffest width (fig.~\ref{fig:Cross-sectional-widths}),
leading to the three order of magnitude difference between the two
curvatures.%
} 

\begin{figure}
\includegraphics[width=3.25in]{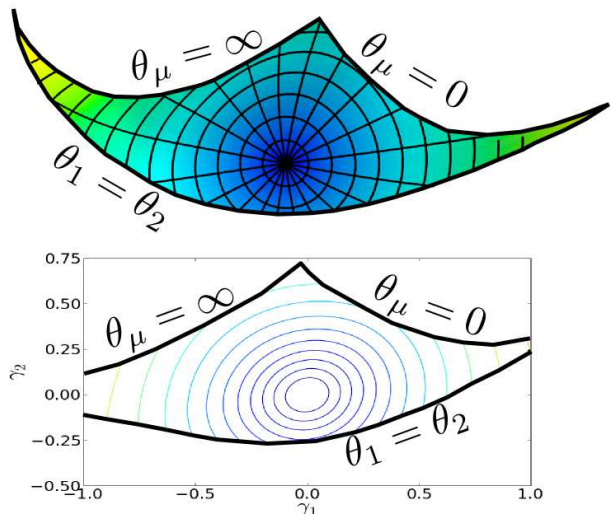} 

\caption{\label{fig:RNCContours} Geodesics can be used to construct polar
coordinates on $\mathcal{M}$ (above). In these new geodesic coordinates,
the cost contours are nearly perfect, isotropic circles (below).}

\end{figure}

We use geodesics to construct new polar coordinates $\gamma^{\mu}=\gamma^{\mu}(\theta)$
on $\mathcal{M}$ that generalize Riemann normal coordinates~\cite{Misner1973}.
Since geodesics are nearly straight lines in data space on $\mathcal{M}$,
we find that cost contours in these coordinates are nearly quadratic
and isotropic around the best fit, as explicitly computed in Fig.~\ref{fig:RNCContours}.
Nonlinear models locally look like linear models with badly chosen
parameters, well beyond the harmonic approximation.

Can these nearly straight geodesics inspire algorithms that lead efficiently
to the best fit? Integrating the geodesic equation with a single Euler
step reproduces the Gauss-Newton step ($\delta\theta^{\mu}=-g^{\mu\nu}\nabla_{\nu}C$
in our notation), ineffective due to the large eigenvalues in the
inverse metric $g^{\mu\nu}$; even geodesic motion hits the boundaries
of $\mathcal{M}$. The ribbon is nearly flat, but very thin; the geodesics
hit the edges long before finding a good fit.

To improve convergence, we can modify the model manifold to remove
the boundaries. One method of doing this is to introduce the {\em
model graph} $\mathcal{G}$, which is the $N$-dimensional parametric
surface drawn by the model embedded in data space crossed with parameter
space. Since most boundaries occur at infinite parameter values, the
model graph $\mathcal{G}$ `stretches' these boundaries to infinity
in the parameter space portion of the embedding. The metric for the
model graph is an interpolation of the data space and parameter space
metrics, $g_{\mu\nu}=g_{\mu\nu}^{0}+\lambda^{*}I_{\mu\nu}$, where
$\lambda^{*}$ determines the weight of the two spaces. Notice that
the eigendirections of the metric are the same on both $\mathcal{M}$
and $\mathcal{G}$; however, the eigenvalues on the graph are given
by $\lambda_{\mathcal{G}}=\lambda_{\mathcal{M}}+\lambda^{*}$. Therefore,
the degenerate eigendirections with $\lambda_{\mathcal{M}}\ll\lambda^{*}$
have eigenvalues $\lambda_{\mathcal{G}}\approx\lambda^{*}$ on the
model graph; $\lambda^{*}$ cuts off the small eigenvalues of the
Hessian. The analogy of the Gauss-Newton step on the model graph is
the well-known Levenberg-Marquardt step, $\delta\theta=-\left(J^{T}J+\lambda^{*}I\right)^{-1}\nabla C$
in our notation~\cite{Marquardt1963,More1977,Press2007}. By dynamically
adjusting $\lambda^{*}$, the algorithm can shorten its step, removing
the danger of the degenerate Hessian, while rotating from the Gauss-Newton
direction into the steepest descents direction. Geometrically we understand
the superiority of Levenberg-Marquardt to be due to the lack of boundaries
of the model graph.

\begin{table}
\centering{}\begin{tabular}{|c||c|c|c|}
\hline 
Algorithm  & Success Rate  & Mean njev  & Mean nfev\tabularnewline
\hline
\hline 
Traditional LM + accel  & 65\%  & 258  & 1494\tabularnewline
\hline 
Traditional LM  & 33\%  & 2002  & 4003\tabularnewline
\hline 
Trust Region LM  & 12\%  & 1517  & 1649\tabularnewline
\hline 
BFGS  & 8\%  & 5363  & 5365\tabularnewline
\hline
\end{tabular}

\caption{\label{tab:Results}\textbf{ }The results of several algorithms applied
to a test problem of fitting a sum of four exponential terms (varying
both rates and amplitudes) in log-parameters (to enforce positivity).
Initial conditions are chosen near a manifold boundary with a best
fit of zero cost near the center of the manifold. Among successful
attempts, we further compare the average number of Jacobian and function
evaluations needed to arrive at the fit. Success rate indicates an
algorithm's ability to avoid the manifold boundaries (find the canyon
from the plateau), while the number of Jacobian and function evaluations
indicate how efficiently it can follow the canyon to the best fit.
BFGS is a quasi newton scalar minimizer of Broyden, Fletcher, Goldfarb,
and Shanno (BFGS) \cite{Nocedal1999,Jones2001}. The traditional \cite{Marquardt1963,Press2007}
and trust region \cite{More1977} implementations of Levenberg-Marquardt
consistently outperform this and other general optimization routines
on least squares problems, such as Powell, simplex, and conjugate
gradient. Including the geodesic acceleration on a standard variant
of Levenberg-Marquardt dramatically increases the success rate while
decreasing the computation time. }

\end{table}

Inspired by the results in Fig.~\ref{fig:RNCContours}, we further
improve the standard Levenberg-Marquardt algorithm. Interpreting the
Levenberg-Marquardt step as a velocity, $v^{\mu}=-g^{\mu\nu}\nabla_{\nu}C$,
where $g$ is the metric on the model graph, the geodesic acceleration
(giving the parameter-effects curvature) is given by $a^{\mu}=-g^{\mu\nu}\partial_{\nu}\vec{y}\cdot\partial_{\alpha}\partial_{\beta}\vec{y}\ v^{\alpha}v^{\beta}$,
giving a step $\delta\theta^{\mu}=v^{\mu}+\frac{1}{2}a^{\mu}$. The
geodesic acceleration is very cheap to calculate, requiring only a
directional second derivative, which can be estimated from three (cheap)
function evaluations (one or two additional function evaluations)
at each step with no extra (expensive) Jacobians. The geodesic acceleration
serves two purposes. First, it provides an estimate for the trust
region in which the linearization approximation (from which Levenberg-Marquardt
is traditionally derived) is valid. At each step, we adjust $\lambda$until
the acceleration is smaller than the velocity, which we find is more
effective at avoiding model boundaries than either tuning until a
downhill step is found~\cite{Marquardt1963,Press2007} or considering
the reduction ratio~\cite{More1977}. The second benefit of the acceleration
occurs when the algorithm must follow a long narrow canyon to the
best fit. In these scenarios convergence may be sped up by approximating
the path with a parabola instead of a line. The utility of the geodesic
acceleration is seen in Table \ref{tab:Results}, where the performance
of several algorithms on a test problem is summarized. More extensive
comparisons and further refined algorithms are in preparation \cite{Transtrum2010b}.

Just as the special sum of squares form of the cost function gives
an approximate Hessian using only first derivatives of the residuals,
$H\approx J^{T}J$, it has (together with the low extrinsic curvature)
allowed us to approximate the {\em cubic} correction using only
a directional second derivative. All other algorithms that seek to
improve the Levenberg-Marquardt algorithm use second derivative information
only to calculate the correction $\delta H_{\mu\nu}=\vec{r}\cdot\partial_{\mu}\partial_{\nu}\vec{r}$,
to the Hessian~\cite{DennisJr1977,Gill1978,DennisJr1981,Gonin1987}.
This correction is negligible if the nonlinearities are primarily
parameter-effects curvature; since the unfit data is nearly perpendicular
to the surface of the model manifold while the nonlinearities are
tangent to the model manifold, the dot product vanishes. Qualitatively,
this means that the approximate Hessian is very accurate and that
the bending of the local ellipses, (due to the third order terms in
the cost that we consider) are the most important correction.

By interpreting the fitting process as a generalized interpolation
scheme, we have seen that the difficulties in fitting are due to the
narrow boundaries on the model manifold, $\mathcal{M}$. These boundaries
form a hierarchy of widths dual to the hierarchy of Hessian eigenvalues
characteristic of nonlinear model fits. Additionally, we both observe
and argue that the model manifold is remarkably flat (low extrinsic
curvature), which leads us to the use of geodesics in the fitting
process. The modified Gauss-Newton and Levenberg-Marquardt algorithms
are understood to be Euler approximations to the geodesic equation
on the model manifold and model graph respectively. The geodesic acceleration
improves convergence of Levenberg-Marquardt by providing a more accurate
trust region while reducing computation time. Data fits are both practically
important and theoretically elegant.

The authors thank Bryan Daniels, Stefanos Papanikolaou, Joshua Waterfall,
Chris Myers, Ryan Gutenkunst, John Guckenheimer, and Eric Siggia for
helpful discussions. This work was supported by NSF grant number DMR-0705167.

\bibliographystyle{aps}
\bibliography{References}

\end{document}